\documentclass[twoside]{dis07}
\usepackage[latin1]{inputenc}
\usepackage[dvips]{graphicx,epsfig,color}
\usepackage{wrapfig,rotating}
\usepackage{amssymb,amsmath,array}

\begin{document}
\title{Search for Excited Leptons at HERA}

\author{TRINH Thi Nguyet\\
On behalf of the H1 Collaboration\\
\vspace{.3cm}\\
%
Centre de Physique des Particules de Marseille\\
163 Avenue de Luminy, F-13288 Marseille Cedex 9, France
}

\maketitle

\begin{abstract}
Searches for excited electrons and neutrinos have been performed using the complete HERA I and II data samples collected by the H1 detector at $\sqrt{s}=320$~GeV corresponding to an integrated luminosity of up to 435~pb$^{-1}$. In absence of a signal, the limits on the ratio of the coupling to the compositeness scale derived extend the excluded region to higher masses than has been possible in previous searches.
\end{abstract}

\section{Introduction}

Compositeness models~\cite{Baur:1990} attempt to explain the hierarchy of masses in the Standard Model (SM) by the existence of a substructure within the fermions. Several of these models predict excited states of the known leptons. Excited leptons ($F^{*}$) are assumed to have the same electroweak SU(2) and U(1) gauge couplings, $g$ and $g'$, to the vector bosons, but are expected to be grouped into both left- and right-handed weak isodoublets with vector couplings. The existence of the right-handed doublets is required to protect the ordinary light leptons from radiatively acquiring a large anomalous magnetic moment via the $F^{*}FV$ interaction (where V is a $\gamma$, $Z$, or $W$). Considering only the electroweak interaction, the phenomenological model describes this interaction by the Lagrangian density:
$$
L_{F^{*}F} = \frac{1}{2\Lambda}{\bar{F^{*}_{R}}}{{\sigma}^{\mu\nu}}[gf\frac{\vec{\tau}}{2}{\partial}_{\mu}{\vec{W_{\nu}}}+g'f'\frac{Y}{2}{\partial}_{\mu}B_{\nu}]{F_{L}} + h.c.
$$
where the new weights $f$ and $f'$ multiply the standard coupling constants $g$ and $g'$ corresponding to the weak SU(2) and the electromagnetic U(1) sectors respectively. The matrix ${{\sigma}^{\mu\nu}}$ is the covariant bilinear tensor, $\tau$ are the Pauli matrices, $W_{{\mu}{\nu}}$ and $B_{{\mu}{\nu}}$ represent the fully gauge invariant field tensors, and Y is the weak hypercharge. The parameter $\Lambda$ has units of energy and can be regarded as the compositeness scale. The relative values of $f$ and $f'$ affect the size of the single-production cross section, their detection efficiencies and also the branching ratios of excited leptons.

Excited electrons and neutrinos may be produced in electron(positron)-proton collisions at HERA via $t$-channel $\gamma$($Z$) or $W^{\pm}$ gauge boson exchange. In the case of excited neutrinos, the cross section is much larger in $e^{-}p$~collsions than in $e^{+}p$~collsions due to the favourable valence u-quark and the helicity enhancement, specific to CC-like processes. Therefore the search for excited neutrinos uses only $e^{-}p$~sample data with an integrated luminosity of 184~pb$^{-1}$. In the case of excited electrons, both $e^{-}p$ and $e^{+}p$~collision modes are used, corresponding to total integrated luminosity of 435~pb$^{-1}$.

\section{Data analysis and results}

Excited leptons ($l=e, \nu$) are searched for in the following decay channels:~$l^{*}\rightarrow{l}{\gamma}$, $l^{*}\rightarrow{l}Z$, $l^{*}{\rightarrow}lW$. The final states resulting from the $Z$ or $W$ hadronic decays are taken into account for both excited electrons and neutrinos and the $Z$ or $W$ leptonic decays are taken into account only for excited neutrinos. In the following, the selection criteria are described for the decay channels.
\vspace{.15cm}\\
\underline {\bf The $\nu^{*}\rightarrow{\nu}{\gamma}$ channel}
\vspace{.15cm}\\
Candidate events are selected by requiring missing transverse momentum $P_{T}^{miss}>$~15~GeV, where the photon is identified as an isolated electromagnetic (e.m.)~cluster in the LAr calorimeter within a polar angle of 5$^{\circ}$ to 120$^{\circ}$. The photon candidates measured within the acceptance of the central tracker ($\theta^{\gamma}>$~20$^{\circ}$) are required to have no associated tracks. The neutral current (NC) and charged current (CC) backgrounds are reduced by imposing the longitudinal momentum balance $E-P_{Z}>$~45~GeV for events with photon candidates at lower transverse momentum $P_T^{\gamma}<$~40~GeV and by requiring the virtuality ($Q_{\gamma}^2$) to satisfy log($Q_{\gamma}^2$)$>$~3.5~GeV$^2$. The background is further suppressed by rejecting events with a transverse momentum of the final hadronic in the calorimeter $P_T^h<$~5~GeV.
\vspace{.15cm}\\
\underline {\bf The $e^{*}\rightarrow{e}{\gamma}$ channel}
\vspace{.15cm}\\
Candidate events are selected with two isolated e.m.~clusters in the LAr calorimeter of transverse energy greater than 20~GeV and 15~GeV, respectively, and with a polar angle between 5$^{\circ}$ and 130$^{\circ}$. The sum of the energies of the two clusters has to be greater than 100~GeV. The background from NC is further suppressed by rejecting events with a total transverse energy of the two isolated e.m.~clusters lower than 75~GeV.
\vspace{.15cm}\\
\underline {\bf The $e^{*}\rightarrow{e}Z,{\nu}W$ and 
$\nu^{*}\rightarrow{e}W,{\nu}Z$  channels with $Z,W\rightarrow{qq'}$ }
\vspace{.15cm}\\
These channels use subsample of events with at least two jets with high transverse momentum $P_T^{j1(j2)}>$~20(15)~GeV reconstructed within 5$^{\circ}<{\theta}^{j1(j2)}<$130$^{\circ}$. The dijet invariant mass must be compatible with the relevant boson mass and should be closest to them.

\begin{table}[]
\begin{center}
\begin{tabular}{|c||c|c||c|}
\multicolumn{4}{c}{Search for e$^*$, $\nu^{*}$ HERA I+II ($\sqrt s = 320$ GeV, preliminary)}\\
\hline
Selection & ~~Data~~ & ~~~~~~~SM~~~~~~~ & Efficiency $\times$ BR \\
\hline
${e}^{*} {\rightarrow} {\nu}{W_{{\hookrightarrow}qq}}$ & $172$ & $175~{\pm}~39$ & $\sim 40$ \% \\
\hline
${e}^{*} {\rightarrow} {e}{Z_{{\hookrightarrow}qq}}$ & $351$ & $318~{\pm}~64$ & $\sim 45$ \%\\
\hline                                        
${e}^{*} {\rightarrow} {e}{\gamma}$ & $112$ & $125~{\pm}~19$ & $60$--$70$ \%\\ 
\hline \hline
${\nu}^{*} {\rightarrow} {\nu}{\gamma}$ & $9$ & $15~{\pm}~4$ & $50$ \%\\
\hline
${\nu}^{*} {\rightarrow} {e}{W_{{\hookrightarrow}qq}}$ & $198$ & $189~{\pm}~33$ &  $30$--$40$ \%\\
\hline
${\nu}^{*} {\rightarrow} {\nu}{Z_{{\hookrightarrow}qq}}$ & $111$ & $102~{\pm}~24$ &  $40$ \%\\
\hline
${\nu}^{*} {\rightarrow} {e}{W_{{\hookrightarrow}\nu\mu}}$  & $0$ & $0.54~{\pm}~0.04$& $3$--$4.5$ \%\\
\hline
${\nu}^{*} {\rightarrow} {e}{W_{{\hookrightarrow}{\nu}e}}$ & $0$ & $0.6~{\pm}~0.3$ & $4$--$6$ \% \\
\hline
 ${\nu}^{*} {\rightarrow} {\nu}{Z_{{\hookrightarrow}ee}}$ & $0$ & $0.12~{\pm}~0.04$ &  $2$ \%\\
\hline
\end{tabular}
\end{center}
\caption{Observed and predicted event yields for the event classes of $e^{*}$ and $\nu^{*}$ searches. The selection efficiency for the signal multiplied by the branching ratio (BR) in each decay channel is also presented.}
\label{tab:estaryields}
\end{table}

{ \bf \it Events with two high $P_T$ jets and one electron}: The channels $e^{*}\rightarrow{e}Z_{\rightarrow{qq}}$ and ${\nu}^{*}\rightarrow{e}W_{\rightarrow{qq}}$ are characterised by at least two high $P_T$ jets and an energetic isolated e.m. cluster $P_T^{e}>$10~GeV ($P_T^{e}>$20~GeV for $e^{*}\rightarrow{e}Z_{\rightarrow{qq}}$) in the polar angle 5$^{\circ}<{\theta}^e<$130$^{\circ}$. The polar angle of the highest $P_T$ jet resulting from $W$ boson should be lower than 80$^{\circ}$. The dijet invariant mass has to be greater than 60~GeV. If $P_T^{e}<$65~GeV, the dijet invariant mass must be greater than 75~GeV. In the case of the ${\nu}^{*}\rightarrow{e}W_{\rightarrow{qq}}$ channel, to reduce the NC background, the polar angle of e.m.~cluster must be lower than 90$^{\circ}$. Furthermore, the background is reduced by requiring the virtuality computed from the e.m.~cluster kinematics $Q^{2}>$2500~GeV$^2$ if $P_T^e<$25~GeV and by requiring a third jet with $P_T>$5~GeV to be present in the event if $P_T^e>$50~GeV.

\begin{figure}[htbp] 
  \begin{center}
    \includegraphics[width=6.8cm]{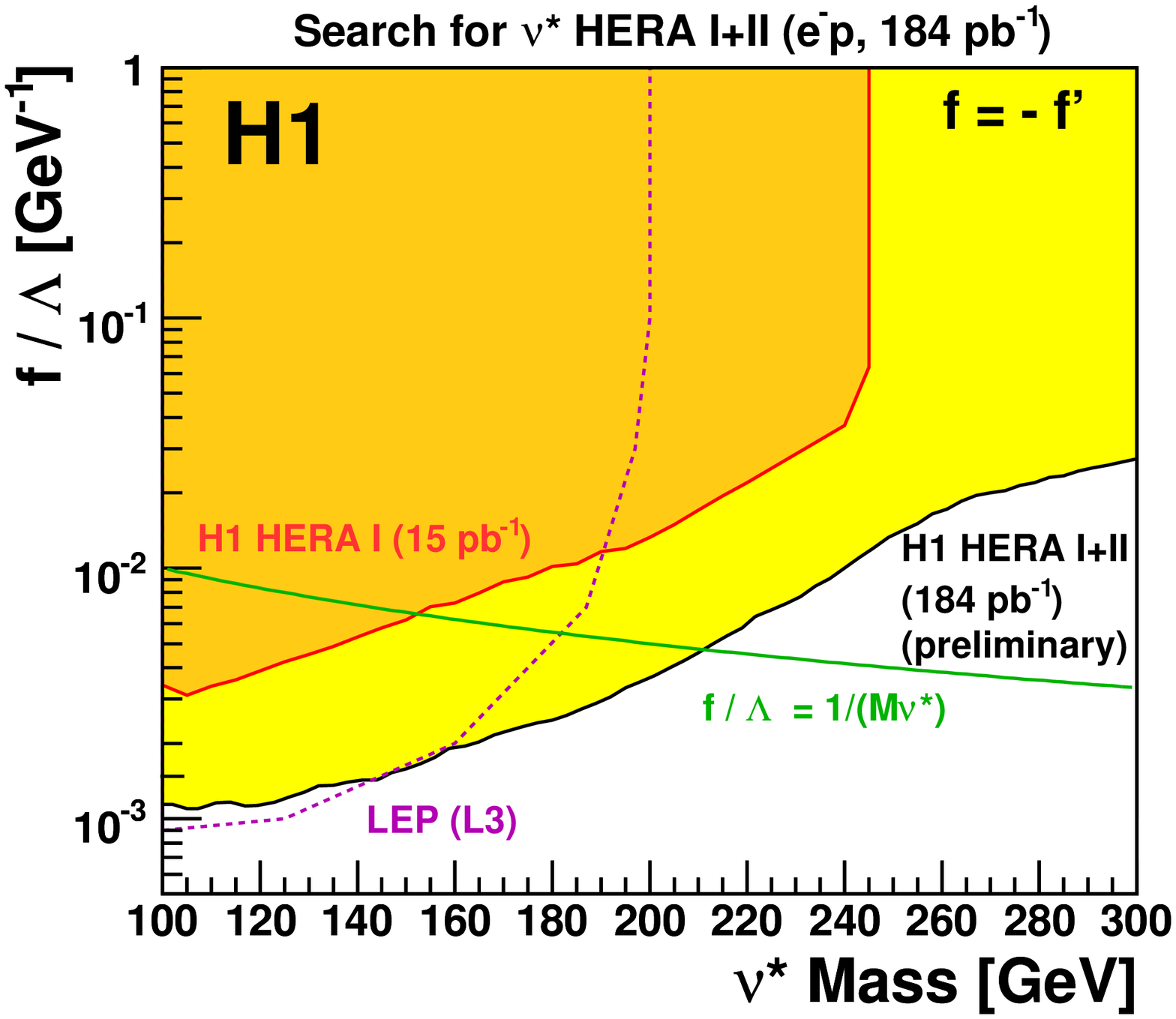}
    \includegraphics[width=6.8cm]{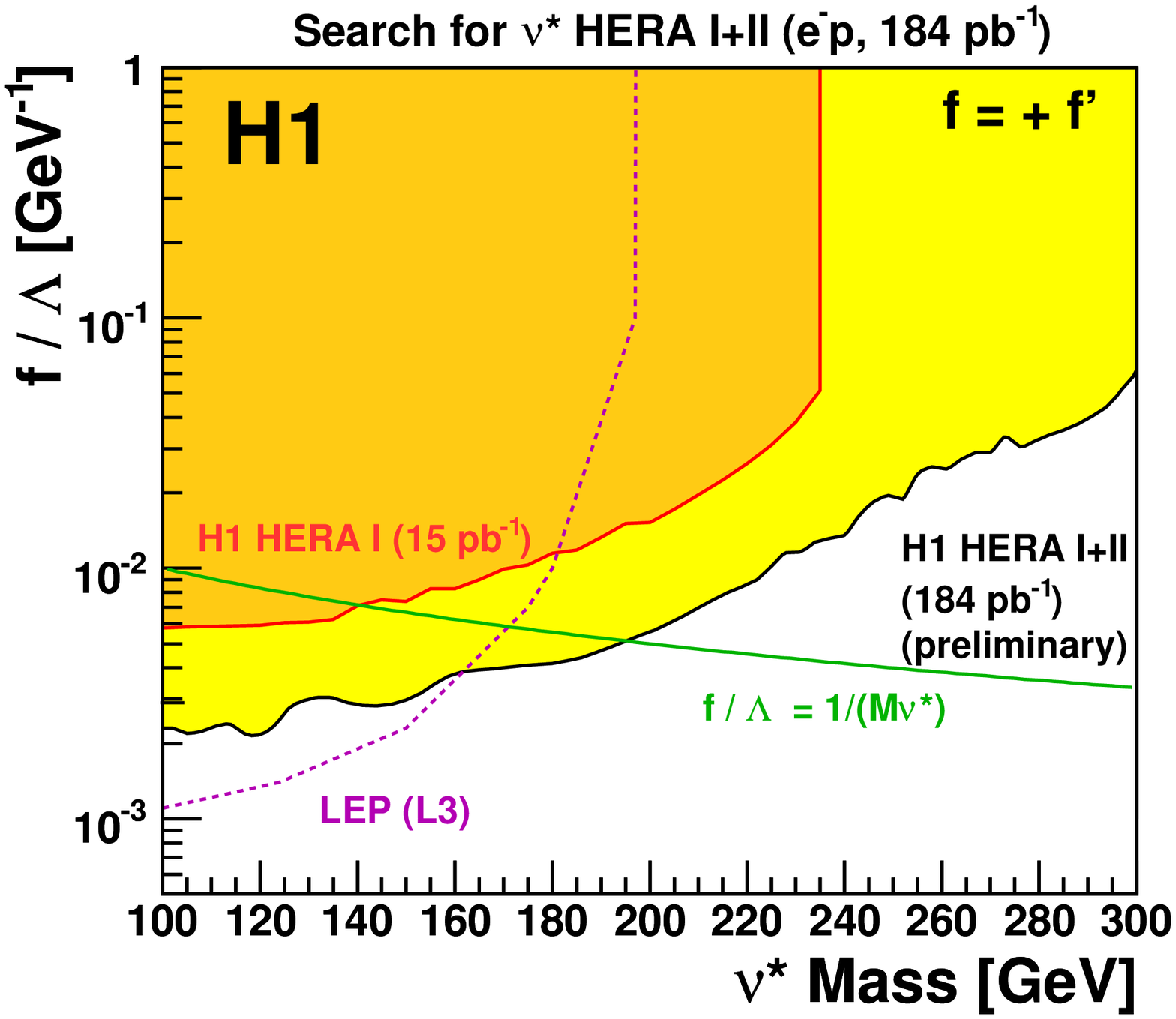}
  \end{center}
  \vspace{-0.5cm}
  \caption{The 95\% C.L.~limits obtained for coupling constants ($f/\Lambda$) as a function of the excited neutrino mass within two assumptions: $f = -f'$~(left) and $f = +f'$~(right). The observed limits from this annalysis using all H1 $e^{-}p$ data is presented by the yellow area. Values of the couplings above the curves are excluded. The orange-dark area corresponds to the exclusion domain published by the H1 experiment using 98/99 data and the dashed line to the exclusion limit from the L3 experiment at LEP~\cite{Acciarri:2000}.}
\label{fig:LimitCouplingNustar}  
\end{figure}

{\bf \it Events with two high $P_T$ jets and $P_T^{miss}$}: The channels $e^{*}\rightarrow{\nu}W_{\rightarrow{qq}}$ and ${\nu}^{*}\rightarrow{\nu}Z_{\rightarrow{qq}}$ are characterised by at least two high jets and $P_T^{miss}>$12~GeV. In the case of $e^{*}\rightarrow{\nu}W_{\rightarrow{qq}}$ channel, the ratio~$V_{ap}/V_p$~of transverse energy flow anti-parallel and parallel to the hadronic final state~\cite{Adloff} is required to be lower than 0.3 to reject the photoproduction (${\gamma}p$) background. The dijet invariant mass is required to be greater than 50~GeV. Furthermore, if $P_T^{miss}<$65~GeV the dijet is required to have an invariant mass above 65~GeV. In the case of the ${\nu}^{*}\rightarrow{\nu}Z_{\rightarrow{qq}}$ channel, a dijet invariant mass greater than 60~GeV is required. In order to reduce CC background, the total hadronic system is required to have a polar angle above 20$^{\circ}$ and a third jet with $P_T>$5~GeV has to be present in the event if $P_T^{miss}<$65~GeV. The longitudinal balance $E-P_Z>$25~GeV is imposed if $P_T^{miss}<$50~GeV. In addition, events with $P_T^{miss}<$30~GeV are only accepted if the topological variable $V_{ap}/V_p>$0.1.
\vspace{.2cm}\\
\underline {\bf The $\nu^{*}\rightarrow{e}W,{\nu}Z$ channels with $Z\rightarrow{ee}$,$W\rightarrow{\nu{e(\mu)}}$}

{\bf \it Events with two electron and $P_T^{miss}$}: These channels use a subsample of events with two high $P_T$ isolated e.m. clusters $P_T^{e1(e2)}>$~20(15)~GeV and a polar angle 5$^{\circ}<{\theta}^{e1(e1)}<$~100$^{\circ}$(120$^{\circ}$) and $P_T^{miss}>$12~GeV. The clusters are required to have associated tracks if they are measured within the acceptance of the central tracker. 

{\bf \it Events with one muon and an electron}: Candidate events containing an islolated muon plus an isolated electron, both having a high transverse momentum $P_T^{e(\mu)}>$~20(10)~GeV and a polar angle 5$^{\circ}<{\theta}^{e(\mu)}<$~100$^{\circ}$(160$^{\circ}$) are selected. A cut $P_T^{miss}>$12~GeV is applied. The backgrounds are reduced by requiring the virtuality ($Q_{e}^2$) to satisfy log($Q_{e}^2$)$>$3~GeV$^2$.  

\vspace{0.3cm}
The observed number of events are compared to the expected SM background in table 1 for each search channel. A good overall agreement is found for all channels. No significant deviation is observed in the data. The selection efficiency for each decay channel for the both $e^*$ and $\nu^{*}$ search is presented also in the table 1. 

\section{Interpretation and Conclusions}

\begin{figure}[htbp] 
  \begin{minipage}{0.25\linewidth}
    \includegraphics[width=6.8cm]{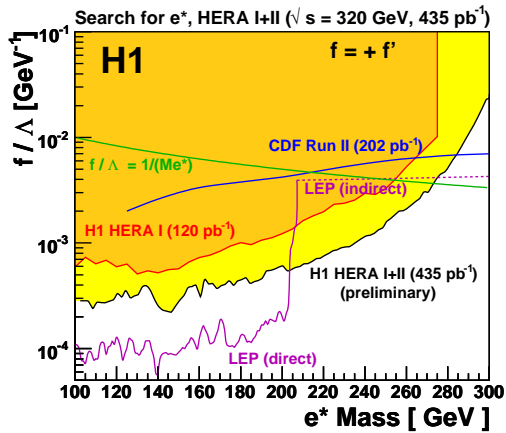}
  \end{minipage}
  \hfill
  \begin{minipage}{0.48\linewidth}
  \caption{The 95\% C.L.~limits obtained for coupling constants ($f/\Lambda$) as a function of the excited electron mass within assumption: $f = +f'$. The observed limit from this analysis using $435~pb^{-1}$~of H1 data is presented by the yellow area. The orange-dark area corresponds to the exclusion domain published by the H1 experiment using HERA I data. The combined exclusion limit from LEP experiments is presented by the violet line. The result of the CDF~\cite{Acosta:2004ri} experiment at the Tevatron is also shown.}
  \end{minipage}
\label{fig:LimitCouplingEstar}  
\end{figure}

In absence of a signal for both excited electron and neutrino production, upper limits on the coupling $f/\Lambda$ have been derived at 95\% Confidence Level~(C.L.) as a function of excited electron and neutrino masses. In case of excited neutrinos production, the obtained limits are displayed for the two assumptions $f=-f'$ and $f=+f'$~(figure 1). For $f=-f'$ (maximal $\gamma{\nu*}{\nu}$ coupling) and assuming $f/\Lambda=1/M_{\nu*}$, excited neutrinos with masses below 211~GeV are excluded at 95\% C.L. The limits on the ratio $f/\Lambda$ also are given for the excited electron for the hypothesis $f=+f'$~(figure 2). We do not consider the case $f=-f'$, because the $\gamma{e^{*}}e$ coupling constant would be equal to zero and the production cross section of the excited electron is very small. For this hypothesis and assuming $f/\Lambda=1/M_{e^{*}}$, excited electrons with masses below 273~GeV are excluded at 95\% C.L.

\begin{footnotesize}

\end{footnotesize}

\end{document}